\documentclass[12pt]{article}
\usepackage{amssymb,epsfig,times,graphicx}
\usepackage{amsmath}
\usepackage{graphicx}
\numberwithin{equation}{section}

\newtheorem{thm}{Theorem}[section]

\newtheorem{prop}[thm]{Proposition}

\newtheorem{defi}[thm]{Definition}

\newtheorem{lemma}[thm]{Lemma}

\newtheorem{rmk}[thm]{Remark}

\newcommand{\beq}{\begin{equation}}
\newcommand{\eeq}{\end{equation}}
\newcommand{\de}{\partial}
\def\d{\partial}
\def\n{\noindent}
\def\f{\frac}

\begin{document}
\title{Natural connections for semi-Hamiltonian systems:\\
The case of the $\epsilon$-system}
\author{Paolo Lorenzoni${}^{*}$, Marco Pedroni${}^{**}$\\
\\
{\small * Dipartimento di Matematica e Applicazioni}\\
{\small Universit\`a di Milano-Bicocca}\\
{\small Via Roberto Cozzi 53, I-20125 Milano, Italy}\\
{\small paolo.lorenzoni@unimib.it}\\
\\
{\small ** Dipartimento di Ingegneria dell'Informazione e Metodi Matematici}\\
{\small Universit\`a di Bergamo}\\
{\small Viale Marconi 5, I-24044 Dalmine BG, Italy}\\
{\small marco.pedroni@unibg.it}}
\date{}

\maketitle

\begin{abstract}
Given a semi-Hamiltonian system, we construct an $F$-manifold with a connection satisfying a suitable compatibility condition with the product. We exemplify this procedure in the case of the so-called
$\epsilon$-system. The corresponding connection turns out to be  flat, and the flat coordinates give rise to additional chains of commuting flows.
\end{abstract}

\section{Introduction}
This paper is a natural continuation of \cite{LPR} and deals with $F$-manifolds
 and their associated integrable hierarchies. The aim of \cite{LPR} essentially was to extend
 the concept of $F$-manifold with compatible connection to the non-flat case and to show its relevance in the theory
 of integrable systems of hydrodynamic type. The integrability condition (i.e., the condition ensuring the existence of integrable flows) in such a framework
  becomes the following simple requirement on the connection $\nabla$ and on the
 structure constants $c^i_{jk}$ entering the definition of these $F$-manifolds:
\beq
\label{compRc}
R^k_{lmi}c^n_{pk}+R^k_{lip}c^n_{mk}+R^k_{lpm}c^n_{ik}=0,
\eeq
where $R^i_{jkl}$ is the Riemann tensor of the connection $\nabla$.
Thus, the starting point in \cite{LPR} was an $F$-manifold, and the goal was the associated integrable hierarchy.
The approach of this paper
 is different, since here we move in the opposite direction.
 The starting point is an integrable hierarchy of hydrodynamic type, and the goal is the construction
 of an $F$-manifold with compatible connection. More precisely, we consider the case of
 {\em semi-Hamiltonian systems\/}, that is, diagonal systems of hydrodynamic type \cite{ts91},
\beq
\label{diag}
u^i_t=v^i(u)\,u^i_x,\qquad u=(u^1,\dots,u^n),\qquad i=1,\dots,n,
\eeq
whose coefficients
$v^i(u)$ (usually called \emph{characteristic velocities}) satisfy the system of equations
\begin{equation}\label{sh}
\partial_j\left(\frac{\partial_k v^i}{v^i-v^k}\right)=
\partial_k\left(\frac{\partial_j v^i}{v^i-v^j}\right)\hspace{1
cm}\forall\, i\ne j\ne k\ne i,
\end{equation}
where $\de_i=\frac{\partial}{\partial u^i}$.
Equations (\ref{sh}) are the integrability conditions for
three different systems:
the first one, given by
\begin{equation}
\label{SYM} \frac{\de_j w^i}{w^i-w^j}=\frac{\de_j v^i}{v^i-v^j},
\end{equation}
provides  the characteristic velocities  of the symmetries
\begin{equation*}
u^i_{\tau}=w^i(u)u^i_x\hspace{1 cm}i=1,...,n,
\end{equation*}
of (\ref{diag}); the second one is the system
\begin{equation}\label{conl}
\d_i\d_j H-\Gamma^i_{ij}\d_i H-\Gamma^j_{ji}\d_j H=0,
\qquad\Gamma^i_{ij}=\f{\de_j v^i}{v^j-v^i},
\end{equation}
whose solutions $H$ are the conserved densities of (\ref{diag});
the third one is
\begin{equation}
\label{ms}
\de_j\ln{\sqrt{g_{ii}}}=\frac{\de_j v^i}{v^j-v^i},
\end{equation}
and relates the characteristic velocities of the system
with a class of diagonal metrics. These metrics and the associated Levi-Civita
 connection play a crucial role \cite{dn89} in the Hamiltonian formalism of (\ref{diag}).

The first result of this paper is the existence of a connection $\nabla$
canonically associated with a given semi-Hamiltonian system, to be called the \emph{natural connection} of the system. By definition, it is compatible---in the sense of (\ref{compRc})---with the product
$c^i_{jk}=\delta^i_j\delta^i_k$, and leads to a structure of $F$-manifold with compatible connection.
We point out that the natural connection \emph{does not coincide},
 in general, with the Levi-Civita connections of the diagonal metrics satisfying the system
(\ref{ms})---whose Christoffel symbols are the $\Gamma^i_{ij}$ appearing in (\ref{conl}).

This alternative approach to the theory of semi-Hamiltonian systems might be
  useful not only from a geometric viewpoint, but also for applications. Indeed, it happens that the connection $\nabla$
 mentioned above turns out to be flat even in cases where the solutions of (\ref{ms}) neither are flat
  nor have the Egorov property\footnote{We recall that a metric is said to have the Egorov property (or to be potential) if there exist coordinates such that the metric is diagonal and $g_{ii}=\d_i\phi$. In other words, the \emph{rotation coefficients}
\beq\label{rotcoef}
\beta_{ij}=\f{\d_i\sqrt{g_{jj}}}{\sqrt{g_{ii}}}
\eeq
must be symmetric}.
 In such a situation the metrics that are compatible
 with the connection $\nabla$ \emph{are not invariant with respect
  to the product}. Nevertheless the flatness of the connection and the condition
$$\nabla_l c^i_{jk}=\nabla_j c^i_{lk},$$
entering the definition of $F$-manifold with compatible connection, are
 sufficient to define the so-called principal hierarchy, as it was shown in \cite{LPR}  
 extending a standard construction of Dubrovin.

As an example, we study
 the semi-Hamiltonian system
\beq\label{epssys}
u^i_t=\left(u^i-\epsilon\sum_{k=1}^n u^k\right)\,u^i_x,\qquad\,\,\,i=1,\dots,n,
\eeq
known in the literature as $\epsilon$-system. It has been studied by several authors.
In particular, we refer to \cite{Fera1,Pv2,tsarev} for the case $\epsilon=-1$, to \cite{FPv} for the case $\epsilon=-\f{1}{2}$, to
\cite{Fera91} for the case $\epsilon=1$,
and to \cite{Pv,plfm,pl} for the general case.
For $n>2$ the metrics
\begin{equation}
\label{sol}
g_{ii}=\f{\varphi_i(u^i)}{\left[\prod_{l\ne i}(u^i-u^l)^2\right]^{\epsilon}},
\hspace{1 cm}i=1,\dots n,
\end{equation}
(where $\varphi_i(u^i)$ are arbitrary non-vanishing functions of a single variable)
 satisfying the system (\ref{ms}) are not of Egorov type. Indeed, their rotation coefficients
\beq\label{rotcoef-eps}
\beta_{ij}=
\left[\f{\prod_{l\ne i}(u^i-u^l)^2}{\prod_{l\ne j}(u^j-u^l)^2}\right]^{\f{\epsilon}{2}}
\f{\epsilon}{u^j-u^i}
\eeq
are not symmetric\footnote{In \cite{pl} it was observed that they satisfy the Egorov-Darboux system
\begin{eqnarray*}
\label{cv1}
&&\d_k\beta_{ij}=\beta_{ik}\beta_{kj}\hspace{1 cm}i\ne j\ne k\\
\label{cv2}
&&\sum_{k}\d_k\beta_{ij}=0\hspace{1 cm}i\ne j\\
\label{qo}
&&\sum_{k}u^k\d_k\beta_{ij}=-\beta_{ij}\hspace{1 cm}i\ne j
\end{eqnarray*}
if the $\varphi_i$ are constant.}. Therefore the natural connection of the $\epsilon$-system does not coincide with the Levi-Civita connection of any of the metrics (\ref{sol}).
 
In the second half of the paper we study in details such natural connection, that turns out to be flat. Thus, besides the usual higher flows of (\ref{epssys}), we obtain $(n-1)$
additional chains of commuting flows associated with the non-trivial flat coordinates of the natural
connection. Remarkably, the recursive identities relating these flows have a double geometrical interpretation: the first one, in terms of the usual recursive procedure involved in the
construction of the principal hierarchy; the second one, in terms of certain cohomological equations appearing in \cite{plfm,pl}.
Finally, we explicitly construct the associated principal hierarchy in the cases $n=2$ and $n=3$.

\subsection*{Acknowledgments}
We thank Andrea Raimondo for useful discussions; many ideas of the present paper have their origin in \cite{LPR},
written in collaboration  with him.
M.P.\ would like to thank the Department {\em Matematica e
Applicazioni\/} of the Milano-Bicocca University for the hospitality. This work has been partially supported
by the European Community through (ESF) Scientific Programme {\em MISGAM}.
Some of the computations have been performed with Maple Software.

\section{$F$-manifolds with compatible connection and related integrable systems}
In this section we review the most important results  of \cite{LPR}.
 First of all let us recall the main definition of that paper.
\begin{defi}
\label{defi:fmancc}
An \emph{$F$-manifold with compatible connection} is a manifold endowed with an associative commutative multiplicative structure given by a $(1,2)$-tensor field $c$
 and a torsionless connection $\nabla$ satisfying condition
\beq\label{scc}
\nabla_l c^i_{jk}=\nabla_j c^i_{lk}
\eeq
and condition
\beq\label{shc}
R^k_{lmi}c^n_{pk}+R^k_{lip}c^n_{mk}+R^k_{lpm}c^n_{ik}=0,
\eeq
where $R^k_{ijl}=\d_j\Gamma^k_{li}-\d_l\Gamma^k_{ji}+
\Gamma^k_{jm}\Gamma^m_{li}-\Gamma^k_{lm}\Gamma^m_{ji}$ is the Riemann tensor of $\nabla$.
\end{defi}
\begin{defi}
An $F$-manifold with compatible connection is called \emph{semisimple} if
 around any point there exist coordinates $(u^1,\dots,u^n)$---called \emph{canonical coordinates}---such that the structures constants
  become
$$c^i_{jk}=\delta^i_j\delta^i_k.$$  
\end{defi}
On the loop space of a semisimple $F$-manifold with compatible connection,
 one can define a semi-Hamiltonian hierarchy. The flows of this hierarchy
  are
\beq\label{flow1}
u^i_t=c^i_{jk}X^k u^j_x,
\eeq
where the vector fields $X$ satisfy the system
\beq\label{admvf}
c^i_{jm}\nabla_k X^m=c^i_{km}\nabla_j X^m.
\eeq
Using the generalized hodograph method \cite{ts91}, one obtains the general solution of (\ref{flow1}) in implicit form as
\beq\label{gom}
t X^i +x e^i= Y^i,
\eeq
where
$$e=\sum_{i=1}^n\f{\d}{\d u^i}$$
is the unity of the algebra and
$Y$ is an arbitrary solution of (\ref{admvf}). Notice that the above representation of the solution
 in terms of critical points of a family of vector fields---as well as the expressions (\ref{flow1},\ref{admvf})---holds true in any coordinate system.

If the connection $\nabla$ is flat, the definition of $F$-manifold with compatible connection reduces to the following definition, due to Manin \cite{manin}.
\begin{defi}
\label{defi:fmanflatc}
An \emph{$F$-manifold with compatible flat connection} is a manifold endowed with an associative commutative multiplicative structure given by a $(1,2)$-tensor field $c$
 and a torsionless flat connection $\nabla$ satisfying condition (\ref{scc}).
\end{defi}

In flat coordinates, condition \eqref{scc} reads
$$\d_l c^i_{jk}=\d_j c^i_{lk}.$$
This, together with the commutativity of the algebra, implies that
$$c^i_{jk}=\d_j C^i_k=\d_j\d_k C^i.$$
Therefore, condition (\ref{scc}) is equivalent to the local existence
 of a vector field $C$ satisfying, for any pair $(X,Y)$ of flat vector fields,
  the condition \cite{manin}:
$$X\circ Y=[X,[Y,C]],$$  
where $\left(X\circ Y\right)^k=c_{ij}^k X^i Y^j$.

The hierarchy associated with an $F$-manifold with compatible \emph{flat} connection
 is usually called \emph{principal hierarchy}. It can been defined in the
  following way, which is a straightforward generalization of the original definition
   given by Dubrovin in the case of Frobenius manifolds.
First of all, one defines the so-called \emph{primary flows}:
\begin{equation}\label{primflo}
u^i_{t_{(p,0)}}=c^i_{jk}X^k_{(p,0)}u^j_x,
\end{equation}  
where  $(X_{(1,0)},\dots, X_{(n,0)})$ is a basis of flat vector fields. Then,  
starting from these flows, one can define the ``higher flows'' of the hierarchy,
\beq
\label{hiflows}
u^i_{t_{(p,\alpha)}}=c^i_{jk}X^k_{(p,\alpha)}u^j_x,
\eeq
by means of the following recursive relations:
\beq\label{recrel}
\nabla_j X^i_{(p,\alpha)}=c^i_{jk}X^k_{(p,\alpha-1)}.
\eeq

\begin{rmk}
The vector fields obtained by means of the recursive relations (\ref{recrel}) are nothing but
the $z$-coefficients
 of a basis of flat vector fields of the deformed connection \cite{du93}
 defined, for any pair of vector fields $X$ and $Y$, by
$$\tilde{\nabla}_X Y=\nabla_X Y+z X\circ Y,\qquad z\in\mathbb{C}\ .$$
\end{rmk}

In the following section we will show how to construct a semisimple $F$-manifold with compatible connection
starting from a semi-Hamiltonian system.

\section{From semi-Hamiltonian system to $F$-manifolds with compatible
 connection: The natural connection}

Let
\beq\label{ssh}
u^i_t=v^i(u)\,u^i_x,\qquad u=(u^1,\dots,u^n),\qquad i=1,\dots,n,
\eeq
be a semi-Hamiltonian system, that is, suppose that the
characteristic velocities $v^i(u)$ satisfy (\ref{sh}).
We want to define a semisimple $F$-manifold with compatible connection
  whose associated semi-Hamiltonian system contains (\ref{ssh}).
 First of all, we define the structure constants $c^i_{jk}$ simply assigning
  to the Riemann invariants $u^i$ the role of canonical coordinates. This means
   that in the coordinates $(u^1,\dots,u^n)$ we have
\beq
\label{prodcc}
c^i_{jk}=\delta^i_j\delta^i_k.
\eeq
Once given the structure constants, the definition of the connection
 $\nabla$ is quite rigid, apart from the freedom in the choice of
  the Christoffel symbols $\Gamma^i_{ii}$. Indeed, such a connection must be torsion-free,
$$\Gamma^i_{jk}=\Gamma^i_{kj},$$
and must satisfy condition (\ref{scc}), that in canonical coordinates reduces to
\begin{eqnarray}
&&\Gamma^i_{jk}=0\qquad\mbox{for $i\ne j\ne k\ne i$},\label{conc1}\\
&&\Gamma^i_{jj}=-\Gamma^i_{ji}\qquad\mbox{for $i\ne j$}.\label{conc2}
\end{eqnarray}
Moreover, the space of solutions of (\ref{admvf}) must contain the characteristic
 velocities of the semi-Hamiltonian system we started with. Putting $X^i=v^i$ in (\ref{admvf}), we obtain
\begin{equation}\label{conc3}
\Gamma^i_{ji}=\f{\d_j v^i}{v^j-v^i}\qquad\mbox{for $i\ne j$},
\end{equation}
where the $v^i$ satisfy (\ref{sh}).
The compatibility condition (\ref{shc}) between $c$ and $\nabla$ is now automatically satisfied, thanks to
the following lemmas.

\begin{lemma}
 \label{lem:R-pre-natconn}
The only non-vanishing components of the Riemann tensor of a connection
satisfying conditions (\ref{conc1},\ref{conc2},\ref{conc3}) are
\begin{equation}\label{nvc1}
R^i_{iki}=-R^i_{iik}=\d_k\Gamma^i_{ii}-\d_i\Gamma^i_{ik}
\end{equation}
and
\begin{equation}\label{nvc2}
R^i_{qqi}=-R^i_{qiq}=\d_q\Gamma^i_{qi}-\d_i\Gamma^i_{qq}+\left(\Gamma^i_{iq}\right)^2
+\Gamma^i_{qq}\Gamma^q_{qi}-\sum_{p=1}^n\Gamma^i_{pi}\Gamma^p_{qq}.
\end{equation}
\end{lemma}

\n
\emph{Proof}.
First of all we have
\begin{eqnarray}
\label{rt1}
R^i_{qkl}&=&0\,\,\,\mbox{for distinct indices}\\
\label{rt2}
R^i_{ikl}&=&-R^i_{ilk}=\d_k\Gamma^i_{il}-\d_l\Gamma^i_{ik}=0\,\,\,\mbox{if $i\ne k\ne l\ne i$}\\
\label{rt3}
R^i_{qki}&=&-R^i_{qik}=\d_k\Gamma^i_{iq}+\Gamma^i_{ik}\Gamma^i_{iq}-\Gamma^i_{iq}\Gamma^q_{qk}
-\Gamma^i_{ik}\Gamma^k_{kq}=0\,\,\,\mbox{if $i\ne q\ne k\ne i$}.
\end{eqnarray}
The first identity is a consequence of the vanishing of the Christoffel
 symbols $\Gamma^i_{jk}$ when the three indices are distinct, the second
  one is a consequence of the semi-Hamiltonian property (\ref{sh}), and the third one is a consequence of the identity
(see \cite{ts91})
$$\d_k\Gamma^i_{iq}+\Gamma^i_{ik}\Gamma^i_{iq}-\Gamma^i_{iq}\Gamma^q_{qk}
-\Gamma^i_{ik}\Gamma^k_{kq}=\f{v^k-v^i}{v^q-v^k}\left[\d_q\left(\f{\d_k v^i}{v^k-v^i}\right)-\d_k\left(\f{\d_q v^i}{v^q-v^i}\right)\right]$$
and of the semi-Hamiltonian property. Finally, using (\ref{conc2}) and (\ref{rt3}) one can easily prove that
\begin{equation}
R^i_{qql}=-R^i_{qlq}=-\d_l\Gamma^i_{qq}+\Gamma^i_{qq}\Gamma^q_{ql}-\Gamma^i_{il}\Gamma^i_{qq}
-\Gamma^i_{ll}\Gamma^l_{qq}=0\,\,\,\mbox{if $i\ne q\ne l\ne i$}.
\end{equation}
All the other components vanish apart from (\ref{nvc1},\ref{nvc2}).
\nopagebreak
\unskip\nobreak\hskip2em plus1fill$\Box$\par\smallskip

\begin{lemma}
Condition (\ref{shc}) follows from (\ref{conc1}), (\ref{conc2}), and (\ref{conc3}).
\end{lemma}

\n
\emph{Proof}. In canonical coordinates, condition (\ref{shc}) takes the form
\beq
\label{Rc-can}
R^n_{lmi}\delta_p^n+R^n_{lip}\delta^n_m+R^n_{lpm}\delta^n_i=0.
\eeq
It is clearly satisfied if $n\ne p,m,i$. Since $p,m,i$ appear cyclicly in (\ref{Rc-can}), it is sufficient to prove it for $p=n$, that is,
\beq
\label{Rc-can2}
R^n_{lmi}+R^n_{lin}\delta^n_m+R^n_{lnm}\delta^n_i=0.
\eeq
In turn, this condition needs a check only for $i\ne n$, leading to $R^n_{lmi}+R^n_{lin}\delta^n_m=0$. We end up with $R^n_{lmi}=0$, where $i\ne n$ and $m\ne n$, which is the content of Lemma \ref{lem:R-pre-natconn}.
\nopagebreak
\unskip\nobreak\hskip2em plus1fill$\Box$\par\smallskip

\begin{rmk}
Let us consider a diagonal metric solving the system (\ref{ms}). Its
Levi-Civita connection clearly satisfies (\ref{conc1}) and (\ref{conc3}). It fulfills also (\ref{conc2}) if and only if the metric is potential in the coordinates $(u^1,\dots,u^n)$. Therefore, only in this case one can choose the $\Gamma^i_{ii}$ in such a way that $\nabla$ is such a
Levi-Civita connection. In other words, a connection $\nabla$ satisfying conditions (\ref{conc1},\ref{conc2},\ref{conc3})
does not necessarily coincide with the Levi-Civita connection of a metric solving (\ref{ms}).
A (counter)example is given by the $\epsilon$-system discussed in Section \ref{sec:eps-sys}.
\end{rmk}

A natural way to eliminate the residual freedom in the choice of the Christoffel coefficients $\Gamma^i_{ii}\,(i=1,\dots,n)$ is to impose
 the additional requirement
\beq\label{unitconst}
\nabla e=0,
\eeq
where $e=\sum_{i=1}^n\f{\d}{\d u^i}$
is the unity of the algebra. Indeed, condition (\ref{unitconst}) means that
\beq
\label{sommagamma}
\sum_{k=1}^n \Gamma_{jk}^i=0.
\eeq
If $i\ne j$, it coincides with (\ref{conc2}).
If $i\ne j$, it gives
\beq\label{conc4}
\Gamma^i_{ii}=-\sum_{k\ne i}\Gamma^i_{ik},\,\,\,\qquad i=1,\dots,n.
\eeq

\begin{defi}
\label{defi:natconn}
We call the connection $\nabla$
defined by conditions (\ref{conc1},\ref{conc2},\ref{conc3},\ref{conc4})
 the \emph{natural connection} associated with the semi-Hamiltonian system
(\ref{ssh}).
\end{defi}
By construction, the natural connection and the product (\ref{prodcc}) satisfy Definition
\ref{defi:fmancc} of $F$-manifold with compatible connection, and the semi-Hamiltonian system (\ref{ssh})
is one of the integrable flows associated with this $F$-manifold.

\begin{rmk}
We added condition (\ref{conc4}) in order to associate a unique connection to a given semi-Hamiltonian system. We will see in Section \ref{sec:eps-sys} that this is a very convenient choice for the $\epsilon$-system. Nevertheless, there might be situations where a different condition has to be chosen.
\end{rmk}

We close this section with a remark on the \emph{Euler vector field}
\beq
\label{Eulervf}
E=\sum_{k=1}^n u^k\f{\d}{\d u^k}.
\eeq

\begin{prop}
The Euler vector field $E$ and the unity of the algebra $e$ satisfy the identity
$$\nabla_e E=e,$$
where $\nabla$ is the natural connection.
\end{prop}

\n
\emph{Proof}. We have that
\begin{eqnarray*}
(\nabla_e E)^i &=& e^k \nabla_k E^i=e^k\left(\d_k E^i+\Gamma^i_{kl}E^l\right)=
\sum_{k=1}^n \left(\delta_k^i+\Gamma^i_{kl}E^l\right) \\
&=& 1+\left(\sum_{k=1}^n \Gamma^i_{kl}\right)E^l=1=e^i,
\end{eqnarray*}
where we have used (\ref{sommagamma}).
\nopagebreak
\unskip\nobreak\hskip2em plus1fill$\Box$\par\smallskip

\section{Special recurrence relations for semi-Hamiltonian systems}
\label{sec:recur}

In view of the example of the $\epsilon$-system (to be discussed in the next section),
we recall some results obtained in \cite{pl,plfm}.

Let $M$ be an $n$-dimensional manifold.
A tensor field $L:T{M}\rightarrow T{M}$ of type $(1,1)$ is said to be \emph{torsionless}
if the identity
\begin{eqnarray*}
[LX,LY]-L[LX,Y]-L[X,LY]+L^2[X,Y]=0
\end{eqnarray*}
is verified for any pair of vector fields $X$ and $Y$ on $M$.  
According to the theory of graded derivations of Fr\"{o}licher-Nijenhuis
 \cite{FN}, a torsionless tensor field $L$ of type $(1,1)$ defines a
 differential operator $d_L$ on the Grassmann algebra of differential forms on $M$, satisfying the
fundamental conditions
\begin{eqnarray*}
d\cdot d_L+d_L\cdot d=0,\hspace{2cm}{d_L}^2=0.
\end{eqnarray*}
On functions and 1-forms this derivation is defined by the following
equations:
\begin{eqnarray*}
&&d_L f(X)=df(LX)\qquad\qquad\mbox{(that is, $d_L f=L^*df$)}\\
&&d_L\alpha(X,Y)=Lie_{LX}(\alpha(Y))-Lie_{LY}(\alpha(X))-\alpha([X,Y]_L),
\end{eqnarray*}
where
\begin{eqnarray*}
[X,Y]_L=[LX,Y]+[X,LY]-L[X,Y].
\end{eqnarray*}
For instance, if $L={\rm diag}(f^1(u^1),\dots,f^n(u^n))$, the action of $d_L$
 on functions is given by the formula
\begin{equation*}
d_L g=\sum_{i=1}^n f^i\f{\d g}{\d u^i}du^i.
\end{equation*}
We assume $a:M\to \mathbb R$ to be a function which satisfies
 the cohomological condition
\begin{equation}
\label{ddla}
dd_L a=0,
\end{equation}
and, from now on, that $L={\rm diag}(u^1,\dots,u^n)$. Then,
according to the results of \cite{pl}, to any solution $h=H(u)$ of the
equation
\begin{eqnarray}
\label{ddlh}
&&dd_L h=dh\wedge da\
\end{eqnarray}
we can associate a semi-Hamiltonian hierarchy. Indeed,
 it is easy to prove that the  system of quasilinear PDEs
\begin{equation}
\label{exF}
u^i_t=\left[-\f{\d_i K}{\d_i H}\right]u^i_x,\hspace{1 cm}i=1,\dots,n,
\end{equation}
is semi-Hamiltonian for any solution $h=K(u)$ of the equation (\ref{ddlh}), and that, once fixed $H$,
 the flows associated to  any pair ($K_1$,$K_2$) of solutions of (\ref{ddlh}) commute.
Moreover, there is a recursive procedure to
 obtain solutions of (\ref{ddlh}).

\begin{lemma}[\cite{DM,Ma}]
\label{lemma:recK}
Let $K_0$ be a solution of $(\ref{ddlh})$. Then
the functions $K_\alpha$ recursively defined by
\begin{equation}
\label{rec}
d K_{\alpha+1}=d_L K_{\alpha}-K_\alpha da,\qquad\alpha\ge 0,
\end{equation}
satisfy equation (\ref{ddlh}).
\end{lemma}

Let us illustrate how to apply the previous procedure in the case of the (trivial) solutions
$H=a$ and $K_0=-a$ of equation (\ref{ddlh}).
Using the recursive relations (\ref{rec}), we get
\begin{eqnarray*}
u^i_{t_0} &=& -\f{\d_i K_0}{\d_i a}u^i_x=u^i_x\\
u^i_{t_1} &=& -\f{\d_i K_1}{\d_i a}u^i_x=[u^i-a]u^i_x=[u^i+K_0]u^i_x\\
u^i_{t_2} &=& -\f{\d_i K_2}{\d_i a}u^i_x=[(u^i)^2+K_0 u^i+K_1]u^i_x\\
          &\vdots& \\  
u^i_{t_n} &=& -\f{\d_i K_n}{\d_i a}u^i_x=[(u^i)^n+K_0(u^i)^{n-1}+K_1(u^i)^{n-2}+\dots+K_{n-1}]u^i_x\\
          &\vdots&
\end{eqnarray*}
The choice $H=-K_0=a=\epsilon\,\mbox{Tr}(L)=\epsilon\sum_{i=1}^n u^i$ gives rise to
the $\epsilon$-system discussed in the Introduction and in the following section.

\section{The $\epsilon$-system}
\label{sec:eps-sys}

In this section we exemplify our construction in the case of the $\epsilon$-system
 (\ref{epssys}). In particular, we show that its natural connection is flat, so that we obtain an $F$-manifold with flat compatible connection and its principal hierarchy.

\subsection{The natural connection of the $\epsilon$-system}
According to Definition \ref{defi:natconn}, the natural connection of the $\epsilon$-system is given by
\begin{equation}
\label{nat-conn-eps}
\begin{aligned}
&\Gamma^i_{jk}=0\qquad\mbox{for $i\ne j\ne k\ne i$}\\
&\Gamma^i_{ji}=\f{\epsilon}{u^i-u^j}\qquad\mbox{for $i\ne j$}\\
&\Gamma^i_{jj}=-\Gamma^i_{ji}=\f{\epsilon}{u^j-u^i}\qquad\mbox{for $i\ne j$}\\
&\Gamma^i_{ii}=-\sum_{k\ne i}\Gamma^i_{ik}=-\sum_{k\ne i}\f{\epsilon}{u^i-u^k}\ .
\end{aligned}
\end{equation}

\begin{prop}
The natural connection of the $\epsilon$-system is flat.
\end{prop}

\noindent
\emph{Proof}. We have that
\begin{equation*}
R^i_{iki}=\d_k\Gamma^i_{ii}-\d_i\Gamma^i_{ik}=
\d_k\left(-\sum_{j\ne i}\f{\epsilon}{u^i-u^j}\right)-\d_i\left(\f{\epsilon}{u^i-u^k}\right)=-\f{\epsilon}{(u^i-u^k)^2}+\f{\epsilon}{(u^i-u^k)^2}=0
\end{equation*}
and
\begin{eqnarray*}
R^i_{qqi}&=&\d_q\Gamma^i_{qi}-\d_i\Gamma^i_{qq}+\left(\Gamma^i_{iq}\right)^2
+\Gamma^i_{qq}\Gamma^q_{qi}-\sum_{p=1}^n \Gamma^i_{pi}\Gamma^p_{qq}\\
&=&\d_q\Gamma^i_{qi}-\d_i\Gamma^i_{qq}+\Gamma^i_{iq}(\Gamma^i_{iq}-\Gamma^q_{iq})
-\sum_{p\ne i,q}\Gamma^i_{pi}\Gamma^p_{qq}-\Gamma^i_{ii}\Gamma^i_{qq}-\Gamma^i_{qi}\Gamma^q_{qq}\\
&=&(\d_q+\d_i)\left[\f{\epsilon}{u^i-u^q}\right]+2\f{\epsilon^2}{(u^i-u^q)^2}
+\sum_{p\ne i,q}\f{\epsilon^2}{(u^i-u^p)(u^p-u^q)}\\
&&+\f{\epsilon}{u^i-u^q}
\sum_{p\ne i,q}\left[-\f{\epsilon}{u^i-u^p}+\f{\epsilon}{u^q-u^p}\right]
+\f{\epsilon}{u^i-u^q}\left[-\f{\epsilon}{u^i-u^q}+\f{\epsilon}{u^q-u^i}\right]\\
&=&\sum_{p\ne i,q}\f{\epsilon^2}{(u^i-u^p)(u^p-u^q)}+\f{\epsilon}{u^i-u^q}
\sum_{p\ne i,q}\f{\epsilon(u^i-u^p-u^q+u^p)}{(u^i-u^p)(u^q-u^p)}=0.
\end{eqnarray*}
Due to Lemma \ref{lem:R-pre-natconn}, there are no more components of $R$ to be checked,
so that $\nabla$ is flat.
\nopagebreak
\unskip\nobreak\hskip2em plus1fill$\Box$\par\smallskip

The next proposition is devoted to the relations between the natural
connection (\ref{nat-conn-eps}) and the Euler vector field (\ref{Eulervf}).

\begin{prop}
The covariant derivative of $E$ is given by $\nabla_j E^k=(1-n\epsilon)\delta_j^k+\epsilon$, that is,
\beq
\label{nablaE}
\nabla E=(1-n\epsilon)I+\epsilon e\otimes d({\rm Tr L}),
\eeq
where $I$ is the identity on the tangent bundle and $e$ is the unity.
Therefore, for any vector field $X$,
\beq\label{Eplusae}
\nabla_X\left(E-\epsilon({\rm Tr L})\,e\right)=(1-n\epsilon) X.
\eeq
Moreover, the Euler vector field $E$ is linear in flat coordinates, i.e.,
\beq
\label{nablanablaE}
\nabla\nabla E=0.
\eeq
\end{prop}

\n
\emph{Proof}. Formula (\ref{nablaE}) follows from the very definitions of $\nabla$ and $E$. Then, using the flatness of $e$, we obtain
(\ref{Eplusae}). Finally,
$$
\nabla\nabla E=\nabla\left((1-n\epsilon)I+\epsilon e\otimes d({\rm Tr L})\right)=0
$$
since also $\nabla I=0$ (due to the fact that $\nabla$ is torsionless) and $\nabla\left(d({\rm Tr L})\right)=0$ (as noticed in the next subsection, before Proposition \ref{prop:flatcoos}).
\nopagebreak
\unskip\nobreak\hskip2em plus1fill$\Box$\par\smallskip

We remark that (\ref{nablanablaE}) is one of the property entering the definition of Frobenius manifold.

\subsection{Flat coordinates}
In this subsection we discuss some properties of the flat coordinates of the natural connection (\ref{nat-conn-eps}) of the
$\epsilon$-system.

We have to find a basis of flat exact 1-forms $\theta=\theta_i du^i$,
that is, $n$ independent solutions of the linear system of PDEs
\beq
\label{flatform}
\begin{aligned}
&\d_j \theta_i-\epsilon\f{\theta_i-\theta_j}{u^i-u^j}=
0,\,\qquad\,\,i=1,\dots,n,\, j\ne i\\
&\d_i \theta_i+\epsilon\sum_{k\ne i}\f{\theta_k-\theta_i}{u^k-u^i}=0,\,\qquad\,i=1,\dots,n,
\end{aligned}
\eeq  
which is equivalent to
\beq
\label{flatform2}
\begin{aligned}
&\d_j \theta_i-\epsilon\f{\theta_i-\theta_j}{u^i-u^j}=
0,\,\qquad\,\,i=1,\dots,n,\, j\ne i\\
&\sum_{k=1}^n\d_k \theta_i=0,\,\qquad\,i=1,\dots,n.
\end{aligned}
\eeq  
In particular, we have that
$$
0=\sum_{k=1}^n \d_k \theta_i=\sum_{k=1}^n \d_i \theta_k=\d_i\left(\sum_{k=1}^n \theta_k\right),
$$
showing that $\sum_{k=1}^n \theta_k$ is constant if $\theta=\theta_k du^k$ is flat.

\begin{rmk}
\label{rmk:flat-ddl}
It trivially follows from (\ref{flatform2}) that $f$ is a flat coordinate if and only if
\beq
\begin{aligned}
 \label{flatcoo1}
& (u^i-u^j)\d_j \d_i f-\epsilon(\d_i f-\d_j f)=
0,\,\qquad\,\,i=1,\dots,n,\, j\ne i\\
& \sum_{k=1}^n\d_k \d_i f=0,\,\qquad\,i=1,\dots,n.
\end{aligned}
\eeq
Since
\begin{equation*}
 (dd_L f-d(\epsilon\,{\rm Tr}L)\wedge df)_{ij}=
(u^i-u^j)\d_j \d_i f-\epsilon(\d_i f-\d_j f),
\end{equation*}
any flat coordinate of the natural connection of the $\epsilon$-system solves equation (\ref{ddlh}) with
$a=-\epsilon {\rm Tr}L=-\epsilon \sum_{i=1}^n u^i$.
\end{rmk}

A trivial solution of the system (\ref{flatform2}) is given by $\theta_j=1$ for all $j$, corresponding to the flat 1-form $\theta^{(1)}=\sum_{j=1}^n du^j=df^1_\epsilon$, where
$f^1_\epsilon=\sum_{j=1}^n u^j$. The other flat coordinates can be chosen according to

\begin{prop}
 \label{prop:flatcoos} There exist flat coordinates $(f^1_\epsilon,f^2_\epsilon,\dots,f^n_\epsilon)$ whose partial derivatives $\d_i f^p_\epsilon(u)$ are homogeneous functions of degree $-n\epsilon$ for all $p=2,\dots,n$ and $i=1,\dots,n$. In particular, if $\epsilon\ne\f{1}{n}$ there exist flat coordinates $(f^1_\epsilon,f^2_\epsilon,\dots,f^n_\epsilon)$ such that
$f^p_\epsilon(u)$ is a homogeneous function of degree $(1-n\epsilon)$ for all $p=2,\dots,n$.
\end{prop}

\n
\emph{Proof}.
Suppose that $\phi=\phi_j du^j$ is a flat 1-form. Then $\sum_{j=1}^n \phi_j=c$ constant, and $\theta:=\phi-\frac{c}{n}\theta^{(1)}$ is still a flat form and satisfies $\sum_{j=1}^n \theta_j=0$.
Then equations (\ref{flatform}) entail that
$$
\sum_{j=1}^n u^j\d_j\theta_i=-n\epsilon\theta_i,
$$
so that $\theta_i(u)$ is homogeneous of degree $-n\epsilon$ for all $i$.  This shows that we can always find a basis
$\left(\theta^{(1)},\theta^{(2)},\dots,\theta^{(n)}\right)$ of flat forms such that the components of $\theta^{(p)}$,
for all $p\ge 2$, are homogeneous of degree $-n\epsilon$. Since flat forms are exact, the first assertion is proved.

The second assertion simply follows from the general fact that a function $f(u)$, whose partial derivatives are
homogeneous of degree $r\ne -1$, is (up to an additive constant) homogeneous of degree $(r+1)$. Even though this is well known, we give a proof for the reader's sake. We know that
\begin{equation*}
\sum_{k=1}^n u^k\d_k\left(\d_j f\right)=r \d_j f,
\end{equation*}
therefore we have
$$\d_j\left(\sum_{k=1}^n u^k \d_k f-(r+1)f\right)=0,$$
meaning that
$$\sum_{k=1}^n u^k\d_k f=(r+1)f+c$$
for some constant $c$ that can be eliminated if $r\ne -1$.
\nopagebreak
\unskip\nobreak\hskip2em plus1fill$\Box$\par\smallskip

In the case $n=2$, we have that
$$\theta^{(2)}=(u^1-u^2)^{-2\epsilon}(du^1-du^2),$$
so that the flat coordinates are
\begin{eqnarray}
\label{fc1}
f_{\epsilon}^1&=&u^1+u^2,\qquad f_{\epsilon}^2=(u^1-u^2)^{1-2\epsilon}\qquad
\mbox{if $\epsilon\ne\f{1}{2}$}\\
\label{fc2}
f_{\epsilon}^1&=&u^1+u^2,\qquad f_{\epsilon}^2=\ln{(u^1-u^2)}\qquad\mbox{if $\epsilon=\f{1}{2}$}.
\end{eqnarray}

In the case $n=3$, assuming  $\epsilon\ne\f{1}{3}$ and taking into account
 the homogeneity of the flat coordinates, it is possible to reduce
  the system (\ref{flatcoo1}) to a third order ODE whose solutions can be explicitly
  written in terms of hypergeometric functions $_2F_1(\alpha;\beta;\gamma;z)$ (see the Appendix for more details).
 If $\epsilon\ne\f{1}{3}$ we obtain the flat coordinates (in the domain where $u^3>u^1$)
\begin{eqnarray*}
f_{\epsilon}^1&=&u^1+u^2+u^3\\
f_{\epsilon}^2&=&(1-3\epsilon)(2u^2-u^3-u^1)[(u^3-u^1)(u^1-u^2)^2]^{-\epsilon}\,_2F_1\left(\epsilon;1-\epsilon;1+2\epsilon;\f{u^2-u^3}{u^1-u^3}\right)+\\
&&(1+\epsilon)[(u^3-u^1)(u^1-u^2)^2]^{-\epsilon}(u^1-u^3)\,_2F_1\left(2-\epsilon;\epsilon-1;1+2\epsilon;\f{u^2-u^3}{u^1-u^3}\right)\\
f_{\epsilon}^3&=&(2u^2-u^1-u^3)[(u^3-u^1)(u^3-u^2)^2]^{-\epsilon}\,_2F_1\left(\epsilon;1-\epsilon;1-2\epsilon;\f{u^3-u^2}{u^3-u^1}\right)+\\
&&-[(u^3-u^1)(u^3-u^2)^2]^{-\epsilon}(u^3-u^1)\,_2F_1\left(\epsilon-1;2-\epsilon;1-2\epsilon; \f{u^3-u^2}{u^3-u^1}\right).
\end{eqnarray*}
It turns out that in the case $\epsilon=\f{1}{3}$ the functions  $f_{\epsilon}^2$ and $f_{\epsilon}^3$ reduce to a constant. For integer
 values of the parameter $\epsilon$ one obtains simpler expressions. For instance, in the case $\epsilon=-1$ (up to inessential constant factors) we have
\begin{eqnarray*}
f_{\epsilon}^1&=&u^1+u^2+u^3\\
f_{\epsilon}^2&=&4(u^3-u^1)^3(u^1+u^3-2u^2)\\
f_{\epsilon}^3&=&4(u^3-u^2)^3(u^2+u^3-2u^1),
\end{eqnarray*}
and in the case $\epsilon=2$ we have
\begin{eqnarray*}
f_{\epsilon}^1&=&u^1+u^2+u^3\\
f_{\epsilon}^2&=&\f{4(u^2+u^3-2u^1)}{(u^2-u^1)^3(u^3-u^1)^3}\\
f_{\epsilon}^3&=&\f{4(u^1+u^2-2u^3)}{(u^3-u^2)^3(u^3-u^1)^3}.
\end{eqnarray*}
This concludes the discussion of the case $\epsilon\ne\f{1}{3}$.
We will make later some consideration for the case $\epsilon=\f{1}{3}$.

\begin{rmk}
Given any set $(f_\epsilon^1,\dots,f_\epsilon^n)$ of flat coordinates, the natural connection
$\nabla$ is the Levi-Civita connection of the metric
$\eta=\eta_{ij} df_\epsilon^i\otimes df_\epsilon^j$ for any choice of the invertible symmetric matrix
$(\eta_{ij})$. For $n=2$, choosing $f_\epsilon^1$ and $f_\epsilon^2$ as above, and
$$
(\eta)_{ij}=\begin{pmatrix}
             0 & 1\\
             1 & 0
            \end{pmatrix},
$$
we obtain
\beq\label{spmet}
g_{ii}=\f{(-1)^i 2(2\epsilon-1)}{\left|u^1-u^2\right|^{2\epsilon}},\qquad i=1,2,
\eeq
which is one of the metrics (\ref{sol}). On the contrary---as we have said in the
Introduction---the metrics (\ref{sol}) are not of Egorov type if $n>2$.
This means that for  $n>2$ the natural connection for the $\epsilon$-system does not coincide with the Levi-Civita connection of the metrics (\ref{sol}).
\end{rmk}

\subsection{The structure constants}
Let us discuss in details the case $n=2$. In the coordinates
 (\ref{fc1}) and (\ref{fc2}) the structure constants are given by
\begin{equation*}
c^1_{11}=c^2_{12}=c^2_{21}=\f{1}{2},\quad
c^2_{11}=c^1_{12}=c^2_{12}=c^2_{22}=0,\quad
c^1_{22}=\f{(f_{\epsilon}^2)^{\f{4\epsilon}{1-2\epsilon}}}{2(2\epsilon-1)^2}
\end{equation*}
for $\epsilon\ne\f{1}{2}$ and by
\begin{equation*}
c^1_{11}=c^2_{12}=c^2_{21}=\f{1}{2},\quad
c^2_{11}=c^1_{12}=c^2_{12}=c^2_{22}=0,\quad
c^1_{22}=\f{1}{8f_{\epsilon}^2}
\end{equation*}  
for $\epsilon=\f{1}{2}$. Hence the vector potential $C$ has components
\begin{eqnarray*}
C^1&=&\f{(f_{\epsilon}^2)^{-\f{2}{2\epsilon-1}}}{4(2\epsilon+1)}+\f{1}{4} (f_{\epsilon}^1)^2\\
C^2&=&\f{1}{2}f^1_{\epsilon} f_{\epsilon}^2
\end{eqnarray*}
for $\epsilon\ne\pm\f{1}{2}$ and
\begin{eqnarray*}
C^1&=&\f{1}{16}f^2_{\epsilon}\left(\ln f^2_{\epsilon}-1\right)+\f{1}{4}(f^1_{\epsilon})^2\\
C^2&=&\f{1}{2}f^1_{\epsilon} f_{\epsilon}^2
\end{eqnarray*}  
for $\epsilon=\pm\f{1}{2}$.

It is easy to check that, lowering the index
of the vector potential with the metric (\ref{spmet}), that in flat coordinates is antidiagonal with components
 $g_{12}=g_{21}=1$, we obtain an exact 1-form.
In other words, for $n=2$ we obtain a scalar potential $F$ satisfying WDVV equations:

\begin{eqnarray}
\label{Fpotential1}
F&=&\f{2\epsilon-1}{4(2\epsilon+1)(2\epsilon-3)}(f_{\epsilon}^2)^{-\f{2\epsilon-3}{2\epsilon-1}}+\f{1}{4}(f_{\epsilon}^1)^2\,f_{\epsilon}^2,\qquad\mbox{if $\epsilon\ne\pm\f{1}{2},\f{3}{2}$}\\
\label{Fpotential2}
F&=&\f{1}{16}(f_{\epsilon}^2)^2\ln{f^2_{\epsilon}}-\f{3}{32}(f_{\epsilon}^2)^2+\f{1}{4}(f_{\epsilon}^1)^2\,f_{\epsilon}^2,\qquad\mbox{if $\epsilon=\pm\f{1}{2}$}\\
F&=&\f{1}{16}\ln{f^2_{\epsilon}}+\f{1}{4}(f_{\epsilon}^1)^2\,f_{\epsilon}^2,\qquad\mbox{if $\epsilon=\f{3}{2}$}
\end{eqnarray}

Let us finally consider the case $n=3$, $\epsilon=1$. As flat coordinates we can  choose
\begin{eqnarray*}
f^1&=&u^1+u^2+u^3\\
f^2&=&{\frac {1}{ 2\left(u^1-u^2 \right)  \left( u^3-u^1 \right) }}\\
f^3&=&{\frac {1}{ 2\left( u^2-u^3 \right)  \left(u^1-u^2 \right) }}.
\end{eqnarray*}
In such coordinates the structure constants read:
\begin{eqnarray*}
&& c^1_{11}=\frac{1}{3},\,
c^2_{11}=0,\,
c^3_{11}=0,\,
c^1_{12}=0,\,
c^2_{12}=\frac{1}{3},\,
c^3_{12}=0,\,
c^1_{13}=0,\,
c^2_{13}=0,\,
c^3_{13}=\frac{1}{3}\\
&&c^1_{22}=-\frac{1}{12}\,{\frac {f^3\, \left( 3\,(f^2)^{2}+3\,{
 f^2}\,f^3+(f^3)^{2} \right) }{(f^2)^{3} \left( {
f^2}+f^3 \right) ^{3}}}\\
&&c^2_{22}=-\frac{1}{24}\,{\frac { \left( 10\,(f^2)^{2}+9\,f^2\,{
 f^3}+2\,(f^3)^{2} \right) (f^3)^{2}\sqrt {2}}{ \left( {
 f^2}+f^3 \right) ^{2}f^2\,\sqrt {-f^3\,f^2\,
 \left( f^2+f^3 \right)  \left( f^3+2\,f^2
 \right) ^{2}}}}\\
&& c^3_{22}=-\frac{1}{8}\,{\frac { \left( f^3+2\,f^2 \right) ^{2}{{
 f^3}}^{3}\sqrt {2}}{(f^2)^{2} \left( f^2+f^3
 \right) ^{2}\sqrt {-f^3\,f^2\, \left( f^2+f^3
 \right)  \left( f^3+2\,f^2 \right) ^{2}}}}\\
&&c^1_{23}=\frac{1}{12}\, \left( f^2+f^3 \right) ^{-3}\\
&&c^2_{23}=-\frac{1}{24}\,{\frac {(f^2)^{2} \left( 4\,(f^2)^{2}+12
\,f^2\,f^3+5\,(f^3)^{2} \right) \sqrt {2}}{ \left( {
 f^2}+f^3 \right) ^{2}f^3\,\sqrt {-f^3\,f^2\,
 \left( f^2+f^3 \right)  \left( f^3+2\,f^2
 \right) ^{2}}}}\\
&&c^3_{23}=\frac{1}{24}\,{\frac { \left( 10\,(f^2)^{2}+9\,f^2\,{
 f^3}+2\,(f^3)^{2} \right) (f^3)^{2}\sqrt {2}}{ \left( {
 f^2}+f^3 \right) ^{2}f^2\,\sqrt {-f^3\,f^2\,
 \left( f^2+f^3 \right)  \left( f^3+2\,f^2
 \right) ^{2}}}}\\
&&c^1_{33}=-\frac{1}{12}\,{\frac {f^2\, \left( (f^2)^{2}+3\,{ f^2
}\,f^3+3\,(f^3)^{2} \right) }{(f^3)^{3} \left( f^2
+f^3 \right) ^{3}}}\\
&&c^2_{33}=\frac{1}{8}\,{\frac { \left( f^3+2\,f^2 \right)  \left(
f^2+2\,f^3 \right) (f^2)^{3}\sqrt {2}}{(f^3)^{2}
 \left( f^2+f^3 \right) ^{2}\sqrt {-f^3\,f^2\,
 \left( f^2+f^3 \right)  \left( f^3+2\,f^2
 \right) ^{2}}}}\\
&&c^3_{33}=\frac{1}{24}\,{\frac {(f^2)^{2} \left( 4\,(f^2)^{2}+12\,
f^2\,f^3+5\,(f^3)^{2} \right) \sqrt {2}}{ \left( {
f^2}+f^3 \right) ^{2}f^3\,\sqrt {-f^3\,f^2\,
 \left( f^2+f^3 \right)  \left( f^3+2\,f^2
 \right) ^{2}}}}
\end{eqnarray*}
We do not display the components of the vector potential $C$, since the
 {corresponding expressions are quite cumbersome}.

\subsection{The primary flows}
In order to define the primary flows we need a basis of flat vector fields
$X=X^i\frac{\partial}{\partial u^i}$,
 that is, $n$ independent solutions of the linear system of PDEs
\beq
\begin{aligned}
\label{flatvf}
&\d_j X^i+\epsilon\f{X^i-X^j}{u^i-u^j}=
0,\,\qquad \,i=1,\dots,n,\,j\ne i\\
&\d_i X^i-\epsilon\sum_{k\ne i}\f{X^k-X^i}{u^k-u^i}=0,\,\qquad \,i=1,\dots,n
\end{aligned}
\eeq
which is equivalent to
\begin{eqnarray}
&&\label{E1}\d_j X^i+\epsilon\f{X^i-X^j}{u^i-u^j}=
0,\,\qquad \,i=1,\dots,n,\, j\ne i\\
&&\label{E2}[e,X]=0.
\end{eqnarray}
Comparing (\ref{flatvf}) with (\ref{flatform}), one notices that the components $X^i$ of a flat vector fields for $\epsilon$ are given by the components of a flat 1-form for $-\epsilon$. Therefore, from
Proposition \ref{prop:flatcoos} we have that there always exists a basis of flat vector fields
$\left(X_{(1)}=e,X_{(2)},\dots,X_{(n)}\right)$ such that the components $X_{(p)}^i(u)$, for $p=2,\dots,n$, are homogeneuos functions of degree $n\epsilon$.

In the case $n=2$ we have, for $\epsilon\ne -\frac12$,
\begin{eqnarray*}
df^1_{-\epsilon}&=&du^1+du^2\\
df^2_{-\epsilon}&=&(1+2\epsilon)(u^1-u^2)^{2\epsilon}(du^1-du^2)
\end{eqnarray*}
and therefore
\begin{eqnarray*}
X_{(1,0)}&=&\frac{\partial}{\partial u^1}+\frac{\partial}{\partial u^2}=e\\
X_{(2,0)}&=&(1+2\epsilon)(u^1-u^2)^{2\epsilon}\left(
\frac{\partial}{\partial u^1}-\frac{\partial}{\partial u^2}\right).\\
\end{eqnarray*}
In canonical coordinates the primary flows are thus given by
\begin{eqnarray*}
u^1_{t_{(1,0)}}&=&u^1_x\\
u^2_{t_{(1,0)}}&=&u^2_x
\end{eqnarray*}
and
\begin{eqnarray*}
u^1_{t_{(2,0)}}&=&(1+2\epsilon)(u^1-u^2)^{2\epsilon}\,u^1_x\\
u^2_{t_{(2,0)}}&=&-(1+2\epsilon)(u^1-u^2)^{2\epsilon}\,u^2_x.
\end{eqnarray*}
The case $n=3$, $\epsilon\ne-\f{1}{3}$ can be treated similarly since we know
the flat coordinates.

Let us consider the case $n=3$, $\epsilon=-\f{1}{3}$. One of the flat vector fields is the unity $e$
of the algebra. We know that there exist two other flat vector
fields $X_{(2)}$ and $X_{(3)}$,
whose components are homogeneous functions of degree -1,
\beq\label{hompr}
u^1\d_1 X_{(i)}^k+u^2\d_2 X_{(i)}^k+u^3\d_3 X_{(i)}^k=-X_{(i)}^k,\qquad  i=2,3,\quad k=1,2,3,
\eeq
satisfying the additional property
\beq\label{sommazero}
X_{(i)}^1+X_{(i)}^2+X_{(i)}^3=0,\qquad i=2,3.
\eeq
Since $\d_j X_{(i)}^k=\d_k X_{(i)}^j$, from (\ref{hompr}) we obtain
\beq\label{hompr2}
u^1 X_{(i)}^1+u^2 X_{(i)}^2+u^3 X_{(i)}^3=c,
\eeq
where $c$ is a constant. Two cases are possible: $c=0$ and $c\ne 0$. In both cases, taking into account condition
(\ref{sommazero}),
we can write one of the components of the vector field $X_{(i)}$ in terms of the remaining two. Substituting the result in
(\ref{flatvf}), we obtain
 a system of $3$ equations whose solution is
\begin{eqnarray*}
X^1_{(2)}&=&\f{(u^2-u^3)^{1/3}}{(u^3-u^1)^{2/3}(u^1-u^2)^{2/3}}\\  
X^2_{(2)}&=&\f{(u^3-u^1)^{1/3}}{(u^2-u^3)^{2/3}(u^1-u^2)^{2/3}}\\
X^3_{(2)}&=&\f{(u^1-u^2)^{1/3}}{(u^3-u^1)^{2/3}(u^2-u^3)^{2/3}}
\end{eqnarray*}
for $c=0$ and
\begin{eqnarray*}
X^1_{(3)}&=&\f{c}{u^2-u^1}+\f{c(u^3-u^2)^{1/3}}{3(u^3-u^1)^{2/3}(u^1-u^2)}\int\f{du^3}{(u^3-u^2)^{1/3}(u^3-u^1)^{1/3}}\\  
X^2_{(3)}&=&\f{c}{u^2-u^1}+\f{c(u^3-u^1)^{2/3}}{3(u^3-u^2)^{2/3}(u^1-u^2)}\int\f{du^3}{(u^3-u^1)(u^3-u^2)^{1/3}}\\
X^3_{(3)}&=&\f{c}{3(u^3-u^2)^{2/3}(u^3-u^1)^{2/3}}\int\f{du^3}{(u^3-u^2)^{1/3}(u^3-u^1)^{1/3}}
\end{eqnarray*}  
for $c\ne 0$. Notice that we can choose the constants of integration in the above integrals in such a way that the $X^i_{(3)}$ be homogeneous of degree -1.

Hence we can explicitly construct the principal hierarchy (\ref{hiflows}) also in the case $\epsilon=-\f{1}{3}$.

\subsection{The higher flows}
The higher flows are defined by vector fields $X_{(p,\alpha)}$ satisfying
\beq\label{recrelbis}
\nabla_j X^i_{(p,\alpha)}=c^j_{ik}X^k_{(p,\alpha-1)}
\eeq
or, more explicitly,
\begin{eqnarray}
\label{HF1}&&\d_j X^i_{(p,\alpha)}+\epsilon\f{X^i_{(p,\alpha)}-X^j_{(p,\alpha)}}{u^i-u^j}=
0,\,\qquad\,i=1,\dots,n,\,j\ne i\\
\label{HF2}&&\d_i X^i_{(p,\alpha)}-\epsilon\sum_{k\ne i}\f{X^k_{(p,\alpha)}-X^i_{(p,\alpha)}}{u^k-u^i}=X^i_{(p,\alpha-1)},\,\qquad\,i=1,\dots,n.
\end{eqnarray}
Taking into account (\ref{HF1}), condition (\ref{HF2}) can be written as
\beq\label{HF2second}
\sum_{k=1}^n \d_k X^i_{(p,\alpha)}=X^i_{(p,\alpha-1)},\,\qquad\,i=1,\dots,n
\eeq
or, in compact form, as
\beq\label{HF2third}
[e,X_{(p,\alpha)}]=X_{(p,\alpha-1)}.
\eeq
We show now that---apart from some critical values of $\epsilon$---the higher flows of the principal hierarchy can be obtained
by applying the recursive procedure described in Section \ref{sec:recur}. First of all, we recall from Remark
\ref{rmk:flat-ddl} that the flat coordinates of the natural connection of the $(-\epsilon)$-system satisfy
equation (\ref{ddlh}), with $L={\rm diag}(u^1,\dots,u^n)$ and $a=\epsilon\mbox{Tr}\, L$.
Therefore, they can be used as starting point for the recursive procedure (\ref{rec}), giving rise to the flows (\ref{exF}),
with $H=a$.

\begin{prop}
Suppose that $\left(f^1_{-\epsilon}={\rm Tr}L,f^2_{-\epsilon},\dots,f^n_{-\epsilon}\right)$
be the flat coordinates described in Proposition \ref{prop:flatcoos} of the natural connection of the $(-\epsilon)$-system.
If $K_{(p,\alpha)}$ are the functions defined recursively by
\beq\label{recak}
K_{(p,0)}=-\epsilon f^p_{-\epsilon},\,\,\,dK_{(p,\alpha+1)}=d_L K_{(p,\alpha)}-\epsilon K_{(p,\alpha)} d({\rm Tr}L),
\qquad\,\,\,\alpha\ge 0,
\eeq
and
\begin{equation}\label{recrel2}
Y_{(p,\alpha)}^i=-\f{\d_i K_{(p,\alpha)}}{\d_i a}=-\frac1{\epsilon}\d_i K_{(p,\alpha)},
\qquad \,\,\,\alpha\ge 0,
\end{equation}
are the components of the vector fields of the corresponding hierarchy,
then the vector fields $X_{(1,\alpha)}=\frac1{\prod_{j=1}^\alpha(j-n\epsilon)}Y_{(1,\alpha)}$ (for $\epsilon\ne \frac{j}{n}$ with
$j=1,\dots,\alpha$) and  $X_{(p,\alpha)}=\frac1{\alpha!}Y_{(p,\alpha)}$, for $p=2,\dots,n$, satisfy the recursion relations
(\ref{recrelbis}).
\end{prop}

\n
\emph{Proof}. We know that from Lemma \ref{lemma:recK} that the function $K_{(p,\alpha)}$ satisfies equation (\ref{ddlh}). Then it is easily checked that the vector fields $Y_{(p,\alpha)}$ and $X_{(p,\alpha)}$ satisfy equation (\ref{HF1}), so that there are only relations (\ref{HF2second}) to be proved.

Let us consider the case $p=1$. After writing (\ref{recak}) in canonical coordinates,
\beq
\label{recakcoo}
\d_j K_{(1,\alpha+1)}=u^j \d_j K_{(1,\alpha)}-\epsilon K_{(1,\alpha)},\qquad\,\,\,\alpha\ge 0,
\eeq
 and recalling that $K_{(1,0)}(u)=-\epsilon\sum_{i=1}^n u^i$, it is clear that one can show by induction that the partial derivatives $\d_j K_{(1,\alpha)}(u)$ are homogeneous functions of degree $\alpha$, so that $K_{(1,\alpha)}(u)$ is homogeneous of degree $(\alpha+1)$.
Using this fact, again from (\ref{recakcoo}) we have that
\begin{equation*}
\sum_{j=1}^n \d_j K_{(1,\alpha+1)}
=\sum_{j=1}^n \left(u^j \d_j K_{(1,\alpha)}-\epsilon K_{(1,\alpha)}\right)
=(\alpha+1-n\epsilon) K_{(1,\alpha)},
\end{equation*}
so that
\begin{equation}
\label{HF2bis}
\sum_{j=1}^n \d_j Y^i_{(1,\alpha+1)}=(\alpha+1-n\epsilon)Y^i_{(1,\alpha)},\,\qquad\,i=1,\dots,n,
\end{equation}
and relations (\ref{HF2second}) for $X^i_{(1,\alpha)}$ follow.

The case $p=2,\dots,n$ can be treated in the same way. The only difference is that the degree of homogeneity
 of $\d_j K_{(p,\alpha)}$ is $(\alpha+n\epsilon)$, so that $K_{(p,\alpha)}$ is homogeneous of degree $(\alpha+1+n\epsilon)$ if $\alpha\ne -1-n\epsilon$.
\nopagebreak
\unskip\nobreak\hskip2em plus1fill$\Box$\par\smallskip

As a consequence of the above proposition we have that, if $\epsilon\ne\f{k}{n}$
for all $k\in\mathbb{N}$, the flows
\begin{eqnarray}
\label{ph1}
u^i_{t_{(1,\alpha)}}&=&X^i_{(1,\alpha)}\,u^i_x=\f{Y^i_{(1,\alpha)}}{\prod_{j=1}^\alpha(j-n\epsilon)}\,u^i_x=-\f{\d_i K_{(1,\alpha)}}{\epsilon\prod_{j=1}^\alpha(j-n\epsilon)}\,u^i_x,\\
\label{ph2}
u^i_{t_{(p,\alpha)}}&=&X^i_{(p,\alpha)}\,u^i_x=\f{Y^i_{(p,\alpha)}}{\alpha!}\,u^i_x=-\f{\d_i K_{(p,\alpha)}}{\epsilon\,\alpha!}\,u^i_x,\qquad p\ne 1,
\end{eqnarray}
(with $i=1,\dots,n$ and $\alpha\ge 0$) define the principal hierarchy of the $\epsilon$-system.

If $\epsilon=\f{k}{n}$ for some $k\in\mathbb{N}$, all the flows (\ref{ph2}) and the flows
 (\ref{ph1}) with $\alpha=0,\dots,k-1$
 still belong to the principal hierarchy. Even though the latter is well defined, relations (\ref{ph1}) do not make sense for $\alpha\ge k$, since the denominator vanishes. The point is that the vector field $Y_{(1,k)}$ is  flat, as one can immediately check using (\ref{HF2bis}), and its components are homogeneous of degree $k=\epsilon n$. Therefore $Y_{(1,k)}$ is a linear combination (with constant coefficients) of the flat homogeneous vector fields $X_{(2,0)},\dots,X_{(n,0)}$. This means that
  $Y_{(1,\alpha)}$ is, for $\alpha\ge k$, a linear combinations of
  the vector fields  $Y_{(p,\alpha-k)}$, with $p=2,\dots,n$. 
  In order to obtain the missing flows of the principal hierarchy, associated  
  to the vector fields $X_{(1,\alpha)}$ with $\alpha\ge k$,  
   one has to solve the system (\ref{HF1},\ref{HF2}) with $p=1,\alpha\ge k$ and $X_{(1,k-1)}=\frac1{\prod_{j=1}^{k-1}(j-k)}Y_{(1,k-1)}$.

For instance, in the case $\epsilon=\f{1}{2},\,n=2$ one can
immediately check that the vector field
\begin{eqnarray*}
Y^1_{(1,1)}&=&u^1-\epsilon {\rm Tr}L=\f{u^1-u^2}{2}\\
Y^2_{(1,1)}&=&u^2-\epsilon {\rm Tr}L=\f{u^2-u^1}{2}
\end{eqnarray*}
is flat, unlike the vector field
\begin{eqnarray*}
X^1_{(1,1)}&=&\f{1}{2}(u^1-u^2)\ln\left(u^1-u^2\right)+\f{3}{2}u^2-\f{1}{2}u^1+c_1
Y^1_{(1,1)}+c_2\\
X^2_{(1,1)}&=&-\f{1}{2}(u^1-u^2)\ln\left(u^1-u^2\right)+\f{3}{2}u^1-\f{1}{2}u^2+c_1
Y^2_{(1,1)}+c_2
\end{eqnarray*}
($c_1$ and $c_2$ are arbitrary constants), obtained solving the system
\begin{eqnarray*}
&&\d_2 X^1_{(1,1)}+\f{1}{2}\f{X^1_{(1,1)}-X^2_{(1,1)}}{u^1-u^2}=0\\
&&\d_1 X^2_{(1,1)}+\f{1}{2}\f{X^1_{(1,1)}-X^2_{(1,1)}}{u^1-u^2}=0\\
&&\d_1 X^1_{(1,1)}-\f{1}{2}\f{X^1_{(1,1)}-X^2_{(1,1)}}{u^1-u^2}=1\\
&&\d_2 X^2_{(1,1)}-\f{1}{2}\f{X^1_{(1,1)}-X^2_{(1,1)}}{u^1-u^2}=1.
\end{eqnarray*}

To conclude, we observe that, under suitable assumptions, the recursion relations
 (\ref{recak}) can be written in a more explicit form. Indeed, we have the following
\begin{prop}
The recursion relations (\ref{recak}) are algebraically solved by
\beq\label{fid}
K_{(1,\alpha)}=\f{1}{\alpha+1}
\left(\sum_{j=1}^n (u^j)^2\d_j K_{(1,\alpha-1)}-\epsilon\left(\sum_{j=1}u^j\right) K_{(1,\alpha-1)}\right)
\eeq  
and, for $\alpha\ne -1-n\epsilon$, by
\beq\label{fid2}
K_{(p,\alpha)}=\f{1}{\alpha+1+n\epsilon}
\left(\sum_{j=1}^n (u^j)^2\d_j K_{(p,\alpha-1)}-\epsilon\left(\sum_{j=1}u^j\right) K_{(p,\alpha-1)}\right)
,\qquad p=2,\dots,n.
\eeq  
\end{prop}

\n
\emph{Proof}. It suffices to multiply (\ref{recakcoo}) by $u^j$ and to sum over $j$, taking into account the homogeneity of the functions $K_{(p,\alpha)}$.
\nopagebreak
\unskip\nobreak\hskip2em plus1fill$\Box$\par\smallskip


\section{Appendix}
Let us consider the system
\begin{eqnarray}
\label{difff}
&&\d_i f=\theta_i,\\
\label{flatness1}
&&\d_j \theta_i-\epsilon\f{\theta_i-\theta_j}{u^i-u^j}=
0,\,\qquad\,\,i=1,2,3,\, j\ne i\\
\label{flatness2}
&&\theta_1+\theta_2+\theta_3=0,\\
\label{hc}
&&u^1 \theta_1+u^2 \theta_2+u^3 \theta_3=(1-3\epsilon)f,
\end{eqnarray}
providing the homogeneous flat coordinates $f$ for the natural connection
 of the $\epsilon$-system in the case $n=3,\,\epsilon\ne\f{1}{3}$.
Using (\ref{flatness2}) and (\ref{hc}) we can write
 two of the components of $\theta$, for instance $\theta_1$ and $\theta_3$,
  in terms of the remaining one and of the flat coordinate $f$:
\begin{eqnarray}
\label{theta1}
\theta_1&=&\f{(u^3-u^2)\theta_2+(1-3\epsilon)f}{u^1-u^3}\\
\label{theta3}
\theta_3&=&\f{(u^2-u^1)\theta_2-(1-3\epsilon)f}{u^1-u^3}.
\end{eqnarray}
Hence, using (\ref{flatness1}) with $i=2,j=1$, we obtain $f$ in terms
 of $\theta_2$:
\beq\label{f}
f=\f{(u^3-u^2)(u^3-u^1)\d_1\theta_2+\epsilon(2u^3+u^1+u^2)\theta_2}{\epsilon(1-3\epsilon)}.
\eeq
In this way equation (\ref{flatness1}) with $i=2,j=3$ reduces to a PDE
 involving only the unknown function $\theta_2$,
\beq\label{theta2}
(u^2-u^3)\d_3\theta_2+(u^2-u^1)\d_1\theta_2-3\epsilon\theta_2=0,
\eeq
whose solution is given by
\beq\label{theta2sol}
\theta_2 =G(u^2,\nu)(u^1-u^2)^{-3\epsilon}
\eeq
where
$$\nu=\frac {u^3-u^2}{u^1-u^2}$$
and $G(u^2,\nu)$ is an arbitrary function. Substituting (\ref{theta2sol})
 in (\ref{f}) and the result in the equation
$$(u^1-u^3)\d_1\d_3 f+\epsilon(\d_1 f-\d_3 f)=0$$
we obtain the third order ODE
\begin{eqnarray*}
&G'''+\f{(4\nu+5\epsilon\nu-4\epsilon-2)}{\nu(\nu -1)}
G''+\f{(9\epsilon\nu^2-13\epsilon^{2}\nu+2\nu^2-2\nu-2
\epsilon+7\epsilon^{2}\nu^{2}+4\epsilon^{2}-9\epsilon\nu)}{\nu^{2}(\nu-1)^{2}}G'+\f{3\epsilon^{2}(2\epsilon-\epsilon\nu-\nu)}{\nu^{2}(\nu-1)^{2}}G=0,
\end{eqnarray*}
where $G''',G'',G'$ are the derivatives of $G(u^2,\nu)$ with respect to $\nu$.  
The above equation can be explicitly solved in terms of hypergeometric functions:
\begin{eqnarray*}
G&=&G_1(u^2)(\nu-1)^{\epsilon}\nu^{-2\epsilon}+
G_2(u^2)(\nu-1)^{-\epsilon}\nu^{-2\epsilon}
\,_2F_1\left(\epsilon;1-\epsilon;1+2\epsilon;\frac{\nu}{\nu-1}\right)\\
&+&G_3(u^2)(\nu-1)^{-\epsilon}
\,_2F_1\left(\epsilon;1-\epsilon;1-2\epsilon;\frac{\nu}{\nu-1}\right),
\end{eqnarray*}
where $G_1,G_2,G_3$ are arbitrary functions of a single variable. The choice
 $G_2=G_3=0$ gives rise to $f=0$, while the choices ($G_2$=constant, $G_1=G_3=0$)
  and ($G_3$=constant, $G_1=G_2=0$) give rise to the homogeneous flat coordinates  
\begin{eqnarray*}
f_{\epsilon}^2&=&(1-3\epsilon)(2u^2-u^3-u^1)[(u^3-u^1)(u^1-u^2)^2]^{-\epsilon}\,_2F_1\left(\epsilon;1-\epsilon;1+2\epsilon;\f{u^2-u^3}{u^1-u^3}\right)+\\
&&(1+\epsilon)[(u^3-u^1)(u^1-u^2)^2]^{-\epsilon}(u^1-u^3)\,_2F_1\left(2-\epsilon;\epsilon-1;1+2\epsilon;\f{u^2-u^3}{u^1-u^3}\right)\\
f_{\epsilon}^3&=&(2u^2-u^1-u^3)[(u^3-u^1)(u^3-u^2)^2]^{-\epsilon}\,_2F_1\left(\epsilon;1-\epsilon;1-2\epsilon;\f{u^3-u^2}{u^3-u^1}\right)+\\
&&-[(u^3-u^1)(u^3-u^2)^2]^{-\epsilon}(u^3-u^1)\,_2F_1\left(\epsilon-1;2-\epsilon;1-2\epsilon; \f{u^3-u^2}{u^3-u^1}\right).
\end{eqnarray*}  
as one can check by a straightforward computation, substituting in the equations
 (\ref{difff},\ref{flatness1},\ref{flatness2},\ref{hc}).

\end{document}